\documentclass[aps,prc,preprint,tightenlines,showpacs,endfloats,byrevtex,%
nofootinbib,superscriptaddress]{revtex4}

\usepackage{epsfig,bm,feynmf}

\begin{document}



\title{The vector-scalar mixing in nuclear medium and the two quark component of scalar meson from QCD sum rules}


\author{Su Houng Lee}%
\email{suhoung@phya.yonsei.ac.kr}

\affiliation{Institute of Physics and Applied Physics,
Yonsei University, Seoul 120-749, Korea}

\author{Takumi Doi}%
\email{doi@pa.uky.edu}

\affiliation{Department of Physics and Astronomy, University of
Kentucky, Lexington KY 40506 USA }

\affiliation{RIKEN BNL Research Center, Brookhaven National
Laboratory, Upton, New York 11973, USA  }

\begin{abstract}
We derive the QCD sum rules for the vector and scalar meson mixing
in nuclear medium, using a two quark interpolating field for both
mesons.  Modeling the mixing via a nucleon hole contribution with
known coupling constant, the sum rule can be used to determine the
overlap of the interpolating field and the scalar meson.  In the
I=0 channel, we find a stable Borel curve and an overlap that is
about 10\%
of the corresponding value in the
pseudo scalar or vector channel. The sum rule in the I=1 channel
is less reliable but also consistent with a small value for the
overlap. These results suggest that both the $\sigma$ and $a_0$
have a small two quark and thus probably a large tetraquark
components. We  discuss the possibility of observing these scalar
mesons from vector mesons emanating from the nuclear medium.
\end{abstract}

\pacs{12.40.Vv, 13.75.-n, 14.65.-q} \keywords{}

\maketitle


\section{Introduction}

The properties of vector meson immersed in nuclear medium is
expected to change due to partial chiral symmetry restoration
and/or many body nuclear effects.  Expected changes range from the
decrease in the mass\cite{BR91,HL92}, increase in the
width\cite{Rapp95,Rapp99,Friman97,Klingl97,Mosel97}, and
appearance of new structures\cite{Lutz01,Mosel06}, every aspect of
which if measured will lead to a new understanding of strong
interaction.   Several experiments are reporting preliminary
results\cite{Ozawa00,Muto05,Trnka05}, which indeed hints to
nontrivial changes and new structures appearing at nuclear matter.
With further refined experiments and with exclusive measurement of
final state particles, the background could be substantially
reduced and vector meson kinematics controlled.   Then detailed
information on the changes of vector meson properties inside
nuclear matter can be extracted, which will then serves as a basis
for understanding symmetry restoration in QCD,  generation of
hadron masses, and QCD phase transition\cite{Mosel98}.  While the
prospects of looking directly into the nuclear matter through
vector meson seems so exciting, a careful model independent
theoretical analysis on the vector meson properties at nuclear
matter have to be carried out simultaneously before any conclusion
on medium effects can be made.

An important feature of the vector meson when immersed into the
nuclear matter is its coupling to the scalar meson with the same
isospin.  Therefore, the longitudinal mode of $\omega$ meson will
couple to $\sigma$, and that of $\rho$ to $a_0$. The coupling is
possible because the nuclear matter expectation value of operators
with Lorentz index (spin) can be non vanishing. Therefore, in
isospin symmetric nuclear matter,  mesons with different spins can
couple as long as their parity and isospin are identical. The
coupling between the scalar  and the vector meson have been
extensively studied in the Walecka model in the isospin 0
channel\cite{Chin,Saito98}, and in the isospin 1
channel\cite{Gale00,Gale01}.     On the other hand, such mixing
have not been extensively studied further in any other model
independent way.  In this respect, it is important to derive model
independent constraints in QCD that can be applied to constrain
the phenomenological parameters in any model calculation.   Here,
we will construct the QCD sum rule for the vector-scalar
correlation function and derive a constraint equation for their
mixing.  Such procedure have been previously used to constrain the
model parameters in the momentum dependent part of the light
vector meson self energy at nuclear matter\cite{FLK99}.

One caveat in working with the scalar meson is the heated
discussion on its  dominant quark content, which could be a
tetraquark\cite{Jaffe76,Jaffe00}.  In the lattice approaches, most
of the calculations using a two quark interpolating field seem to
predict the ground state mass of scalar particle in the isospin 1
channel to be above 1.3
Gev\cite{Bardeen04,Prelovsek04,Suganuma05,Mathur06,Burch06}, while
some predict it to be around 1 GeV\cite{McNeile06,Hashimoto}.  In
contrast, most lattice calculations based on a four quark
interpolating field current consistently predict the mass to be
around 1 GeV for the $f_0,a_0$\cite{Suganuma05,Ishii06} and around 600
MeV for the $\sigma$\cite{Mathur06}.  Some calculations based on
two quark current also give some consistent prediction for
$\sigma$\cite{Kunihiro03,Wada07}.  So while the lattice
calculations seems to favor a tetraquark picture for the scalar, a
two quark component is not ruled out.  This is in a sense not
surprising as the true scalar should have both a two quark and a
four quark component, and the real question is what their relative
contribution to the total wave function should be.

In the QCD sum rule approach for the scalar meson,  a previous
work based on a two quark interpolating field\cite{Reinders84}, is
known to be controversial\cite{Shuryak93}.  This is so because the
single instanton configuration is expected to contribute to the
correlation functions between the scalar currents or between the
pseudo scalar currents, and spoil the respective convergence of
the operator product expansion (OPE). In the present analysis, we
will investigate the correlation function between the scalar and
the vector current, where both interpolating fields are composed
of a quark and an anti-quark. Thus, the OPE are free from the
single instanton contribution and a reliable QCD sum rule analysis
will be possible.  As we will see, by investigating the scalar
through a two quark current, and modeling the phenomenological
side through a particle hole contribution with known couplings, we
find that for both $a_0(980)$ and $\sigma$ meson, its overlap to
the two quark current is
less than 20\%
of the values
typically expected from the usual meson. Hence, our result
suggests that the scalar nonet have a dominant tetraquark
component\cite{Jaffe76}. Furthermore, our analysis confirms that
the scalar vector mixing indeed takes place in the physical
time-like region.    While the $\sigma$ meson is too wide to be
observable in the $\omega$ meson emanating from nuclear matter,
the $a_0$ will be in the $\rho$.

\section{OPE}

We start from the correlation function between the scalar and
vector meson.
\begin{eqnarray}
\Pi_\mu(q)=i \int d^4 x e^{iqx} \langle T [ J(x),J_\mu(0)] \rangle
\label{corr}
\end{eqnarray}
where $J^{\sigma,a_0}= \frac{1}{2}(\bar{u}u \pm \bar{d}d )$ and
$J^{\omega,\rho}_\mu= \frac{1}{2}(\bar{u}\gamma_\mu u \pm \bar{d}
\gamma_\mu d )$.

In the vacuum, the correlation function is zero as there can not
be any coupling between the scalar and vector current.  However,
in nuclear medium Eq.(\ref{corr}) does not vanish, as the medium
provides the necessary four momentum.

The OPE of Eq.(\ref{corr}) at nuclear matter can be obtained in
the standard way\cite{HL92,HKL93,Lee97}. To leading order in
coupling, the first non-vanishing operator occurs at dimension 6.
Taking the quark operators only, it is given as

\begin{eqnarray}
\Pi^\mu & = & {(g^{\mu \nu}-q^\mu q^\nu/q^2  )\over q^4}
\bigg(-\frac{160 \pi}{81} \alpha_s \bigg) \bigg( \langle \bar{u}
u\rangle_0 \langle \bar{u} \gamma^\mu u \rangle_{n.m.} + \langle
\bar{d} d\rangle_0 \langle \bar{d} \gamma^\mu d \rangle_{n.m}
\bigg) , \label{ope}
\end{eqnarray}
for both the isospin 1 and 0 channel. Here, $\langle \cdot
\rangle_0$ ($\langle \cdot \rangle_{n.m.}$) denotes the vacuum
(nuclear matter) expectation value.    We have assumed vacuum
saturation to extract the vacuum expectation value of the quark
condensates. For a symmetric nuclear matter at rest, the
expectation value becomes to leading order in nuclear density,
\begin{eqnarray}
\bigg( \langle \bar{u} u\rangle_0 \langle \bar{u} \gamma^\mu u
\rangle + \langle \bar{d} d\rangle_0 \langle \bar{d} \gamma^\mu d
\rangle \bigg) \rightarrow  \langle \bar{q} q \rangle_0 \rho_N 3
\delta^{\mu 0},
\end{eqnarray}
where $\rho_N,m_N$ are the symmetric nuclear matter density and
nucleon mass respectively.  Here we have made use of the linear
density approximation $\langle \cdot \rangle_{n.m}=\langle \cdot
\rangle_0+\frac{\rho_N}{2m_N} \langle N| \cdot |N \rangle $.

\section{Phenomenological side}

Consider taking the nuclear matter expectation value of the
correlation function in Eq.(\ref{corr}).  Since this correlator
vanishes in the vacuum, it is just the nucleon expectation value
times the density factor in the linear density approximation. If
we saturate the intermediate states by hadronic states, the
contribution from the ground states becomes as follows
\begin{eqnarray}
\Pi^\mu=\frac{\rho_N}{2 m_N}
\frac{if_S}{q^2-m_S^2}M^\nu_{S+N\rightarrow V+N} \bigg(g^{\nu
\mu}-q^\nu q^\mu/q^2 \bigg)\frac{if_V}{q^2-m_V^2}. \label{phen}
\end{eqnarray}
Here,
\begin{eqnarray}
\langle 0| J|S\rangle & = & f_S \nonumber \\
\langle 0| J^\mu |V\rangle & = & \epsilon^\mu f_V ,
\label{overlap}
\end{eqnarray}
where $S,V$ is the scalar and vector meson, and $\epsilon^\mu$ the
polarization tensor of the vector meson. $M^\nu_{S+N\rightarrow
V+N}$ is the forward scattering matrix element.

Let us assume the following phenomenological Lagrangian,
\begin{eqnarray}
{\cal L}=\bigg( g_V \bar{N} \gamma_\mu \tau^a
N-\frac{\kappa}{2m_N}\bar{N} \sigma_{\mu \nu} \tau^a N
\partial_\nu \bigg) V^\mu + g_S \bar{N} S \tau^a N, \label{lag}
\end{eqnarray}
where we define $\tau^a = \sigma^a$ ($\sigma^a$: Pauli matrices)
for the isospin 1 channel and
$\tau^a = 1$ for the isospin 0 channel.
Using
this, the forward scattering matrix element has the following
form.
\begin{eqnarray}
M^\nu_{S+N\rightarrow V+N} & =& g_V g_S \bigg( g^{\nu \mu}-q^\nu
q^\mu/q^2 \bigg) \frac{4m_N}{q^2-4(p\cdot q)^2/q^2} 2m_N
\bar{N}\gamma_\mu (\tau^a)^2 N
\nonumber \\
&& +  \frac{\kappa_a}{m_N}  g_S \bigg( g^{\nu \mu}-q^\nu q^\mu/q^2
\bigg) \frac{q^2}{q^2-4(p\cdot q)^2/q^2} 2m_N \bar{N}\gamma_\mu
(\tau^a)^2 N  \label{matrix}
\end{eqnarray}
Substituting Eq.(\ref{matrix}) into Eq.(\ref{phen}), one finds
that Eq.(\ref{phen}) can be written in the following form,
\begin{eqnarray}
\Pi^\mu=\Pi(q^2,p\cdot q)^{phen} \times \bigg( g^{\nu \mu}-q^\nu
q^\mu/q^2 \bigg) \bar{N}\gamma_\mu (\tau^a)^2 N. \label{inv-form}
\end{eqnarray}
For the nucleon at rest, the tensor part combined with the nucleon
expectation value is proportional to $|\vec{q}|^2$.  This means
that the  vector scalar coupling vanishes when $\vec{q}
\rightarrow 0$.

The sum rule constraint is obtained from identifying the
$\Pi^{phen}$ in Eq.(\ref{inv-form}) to the corresponding OPE in
Eq.(\ref{ope}).  In the limit where $q=(\omega,0,0,0)$, we find
the following sum rule.

\begin{eqnarray}
\bigg(- \frac{160 \pi }{27} \alpha_s \bigg) \langle \bar{q}
q\rangle_0 \frac{1}{\omega^4} & =&  -\bigg(f_Sf_V g_S g_V
\bigg)\frac{1}{(\omega^2-m_f^2)} \frac{1}{(\omega^2-m_V^2)}
\frac{4
m_N}{(\omega^2-4m_N^2)} \nonumber \\
&& - \bigg(f_Sf_V g_S  \frac{\kappa}{m_N}
\bigg)\frac{1}{(\omega^2-m_f^2)} \frac{1}{(\omega^2-m_V^2)}
\frac{\omega^2}{(\omega^2-4m_N^2)}+ ...., \label{sumrule}
\end{eqnarray}
where  the dots represent contributions from excited meson states.

 The first term in the right hand side of
Eq.(\ref{sumrule}) represents the contributions from the vector
coupling in Eq.(\ref{lag}), while the second that from the tensor
coupling. It should be noted that since we have calculated only
the leading term of the OPE, the QCD sum rule will constrain only
one parameter in the phenomenological side.

\section{sum rules}

\subsection{$\omega-\sigma$ mixing}

Let us start from considering the isospin 0 channel. Here, the
mixing of the $\omega$ will be to the $\sigma$.  As discussed
before, many phenomenological and theoretical considerations lead
us to believe that the dominant quark content of the scalar nonet
is a tetraquark\cite{Jaffe76}.    The previous QCD sum rule
calculation for the isospin 0 and 1 scalar meson with a two quark
interpolating current, seemed to reproduce the masses of $a_0,f_0$
to be of 1 GeV and degenerate\cite{Reinders84}.  However, direct
instanton contribution was found to break the degeneracy, such
that in the isospin 1 channel, the sum rule couple dominantly to
$a_0(1450)$, and in the isospin 0 channel to a state with a mass
smaller than 1 GeV, which could be the
$\sigma(600)$
\cite{Elias98}.
On the other hand, recent
calculation with tetraquark interpolating currents do seem to
reproduce the masses of the scalar nonet correctly, if direct
instanton contributions are taken into
account\cite{Du04,Nielsen04,Hosaka06-1,Leehj06}.

The culprit for the complication in the scalar channel is the
direct contributions from a single instanton\cite{Shuryak93}. In
the present work, since we are considering the non diagonal
correlation function between the scalar and the vector current, no
direct instanton will contribute to the zero-modes.
Moreover, since the ground sate
will dominate the sum rules, we can get direct information on
$f_S$, which is the overlap of the ground state scalar meson to
the two quark interpolating field.

The mixing in the isospin 0 channel is slightly complicated
because the $\sigma$ width is very large and the $\omega$ will
also mix strongly with $f_0(980)$.   On the other hand, we have
calculated only the leading term in the OPE, and consequently can
constrain only one parameter. Therefore,  as a first
approximation, we will take the $\sigma$ width to be small,
neglect the contribution from $f_0(980)$, and assume that the
$\omega$ nucleon coupling is dominated by the vector part such
that  $\kappa=0$, as is phenomenologically motivated.   We
then perform the Borel transformation, which will suppress the
contribution from $f_0(980)$, and then compare the OPE to the
phenomenological side.

After the Borel transformation, the sum rule for the
$\omega-\sigma$ mixing becomes,
\begin{eqnarray}
\bigg(- \frac{160 \pi }{27} \alpha_s \bigg) \langle \bar{q}
q\rangle_0 \frac{1}{M^2}  & =&  \bigg(f_\sigma f_\omega g_\sigma
g_\omega 4m_N\bigg) \bigg[\frac{1}{(m_\sigma^2-m_\omega^2)}
\frac{1}{(m_\sigma^2-4m_N^2)} e^{-m_\sigma^2/M^2} \nonumber \\
&& + \frac{1}{(4m_N^2-m_\sigma^2)} \frac{1}{(4m_N^2- m_\omega^2)}
e^{-4m_N^2/M^2} \nonumber \\
&&+ \frac{1}{(m_\omega^2-4m_N^2)}
\frac{1}{(m_\omega^2-m_\sigma^2)} e^{-m_\omega^2/M^2} \bigg] .
\label{omega}
\end{eqnarray}
The sum rule in Eq.(\ref{omega}) can be used to constrain the
whole coefficient $f_\sigma f_\omega g_\sigma g_\omega$. However,
 phenomenologically $g_\omega, g_\sigma, f_\omega$ are all rather well known.
Therefore,   we will use the sum rule to constrain the parameter
$f_\sigma$.   For the parameters, we take  $g_\omega = g_{NN \omega}=11.5$,
$g_\sigma = g_{NN \sigma}=9.4$\cite{Bonn1},
$m_\sigma=0.55$ GeV,
and $f_\omega=0.12$
GeV$^2$\cite{klingl}.  The same value of $f_\omega$ can be
obtained from the QCD sum rule calculation for the vector
mesons\cite{HKL93} from which we also take  $f_\omega=f_\rho$.
For the parameters appearing in the OPE, we
will take the following values,
\begin{eqnarray}
\langle \bar{q} q \rangle _{\mu={\rm 1 GeV}}& = & -(0.23 {\rm GeV})^3, \nonumber \\
\alpha(M^2) & = & \frac{4\pi}{9 \ln(M^2/(0.2 {\rm GeV})^2)}.
\end{eqnarray}

Fig 1 shows the sum rule result for $f_\sigma$ normalized by
$\frac{1}{2}\langle 0| \bar{q} i \gamma^5 \tau^0 q|\pi^0
\rangle=0.13 $GeV$^2$ calculated in the soft pion limit. One notes
the sum rule has a plateau at Borel mass of around $1.0-2.0$ GeV$^2$, at
which the overlap is about 10\% of that of the pion. This value
fits very well to the picture that the scalar nonet is mainly a
tetraquark and has a very small quark anti-quark component.
Because the parameters for $\sigma$ have some uncertainties, we
check the sensitivity of our result to $m_\sigma$ and $g_\sigma$.
For $m_\sigma$ uncertainty, we reanalyze the Borel plot for
$m_\sigma \simeq 0.4-0.8$GeV, and confirm that the change of the
result is less than 10\%. This seems surprising at first when
looking at the individual terms in the right hand side of
Eq.(\ref{omega}).  But it just reflects the fact that for the
overall sum, the dependence on $m_\sigma$ is weak, as can be seen
from expanding the right hand side of Eq.(\ref{omega}) in $1/M^2$
and noting that the leading term has no dependence in $m_\sigma$.
For the uncertainty of $g_\sigma$, we consider the larger value of
$g_\sigma = 14-17$ which is recently
obtained\cite{nijmegen,erkol}. However, using these values leads
to reducing our result of $f_\sigma$ by about a factor of $1/2$,
which indicates the quark anti-quark component is further
suppressed.

Finally, we comment that
the overlap
 $f_\sigma$ to that of the pion overlap constant ($=0.13$GeV$^2$)
is not special as the corresponding
overlap value for the vector meson is $f_\rho = f_\omega=0.12 $ GeV$^2$ and
$f_{a_1}\simeq 0.17$ GeV$^2$\cite{Reinders84} for the axial vector meson, and
therefore, $f_\sigma$ is much smaller than any of these values.

\subsection{$\rho-a_0$ mixing}

The sum rule for the isospin 1 channel can be obtained similarly.
In this case, however,  the $\rho$ coupling to the nucleon is
dominated by the large tensor coupling and we will assume $g_\rho
= g_{NN \rho}=0$. For the other couplings, we will take
$\kappa=14.2$ and $g_{a_0} = g_{NN a_0}=2.8$ from the Bonn
potential\cite{Bonn1}. For the overlap constant for the vector
meson, we again take $f_\rho=0.12$GeV$^2$\cite{klingl}.   The
Borel sum rule for this case then becomes,

\begin{eqnarray}
\bigg(- \frac{160 \pi }{27} \alpha_s \bigg) \langle \bar{q}
q\rangle_0 \frac{1}{M^2}  & =&  \bigg(f_{a_0} f_\rho g_{a_0}
\kappa /m_N\bigg) \bigg[\frac{m_{a_0}^2}{(m_{a_0}^2-m_\rho^2)}
\frac{1}{(m_{a_0}^2-4m_N^2)} e^{-m_{a_0}^2/M^2} \nonumber \\
&& + \frac{4m_N^2}{(4m_N^2-m_\rho^2)} \frac{1}{(4m_N^2-
m_{a_0}^2)}
e^{-4m_N^2/M^2} \nonumber \\
&&+ \frac{m_\rho^2}{(m_\rho^2-4m_N^2)}
\frac{1}{(m_\rho^2-m_{a_0}^2)} e^{-m_\rho^2/M^2} \bigg] .
\label{rho}
\end{eqnarray}
The sum rule for $f_{a_0}$ can be obtained by dividing the right
hand side of Eq.(\ref{rho}) by its left side apart from $f_{a_0}$.
Fig. \ref{rho-plot} shows the the value for $f_{a_0}$, normalized
by the corresponding pion value. Unfortunately, there is no stable
Borel region from which we can reliably determine the value.
Adding the vector coupling of the $\rho$ will not change much. The
problem in this case, could be due to the nontrivial contribution
from $a_0(1450)$, which is expected to be a dominant quark
anti-quark state.  In fact, previous QCD sum rule calculation with
two quark current in the isospin 1 channel find the ground state
mass to be around 1400 MeV\cite{Hosaka06-2}, suggesting that the
$a_0(1450)$ dominantly contributes to the sum rule. Therefore, in
the sum rule in Eq.(\ref{rho}), while the Borel transformation
suppresses the contribution from the $a_0(1450)$, its contribution
might not be suppressed due to the large overlap $f_{a_0(1450)}$.
Unfortunately, the relevant couplings for $a_0(1450)$ are not
known, and can not be subtracted out from the sum rule to obtain
$f_{a_0}$.

Another reason why the sum rule in Eq.(\ref{rho}) is less reliable
than that in Eq.(\ref{omega}), can be seen from the
phenomenological side.  Apart from the couplings, the right hand
side of Eq.(\ref{rho}) can be obtained by taking $M^4 d/dM^2$ of
the right hand side of Eq.(\ref{omega}).  Such derivatives tend to
enhance the contributions from higher energy states, as additional
factor of resonance mass is multiplied in front of the exponential
suppression factors.  Therefore, while the contribution
proportional to $e^{-m_\sigma^2/M^2}$ is larger than that
proportional to $e^{-m_\omega^2/M^2}$ and that to negligible $e^{-4m_N^2/M^2}$
in Eq.(\ref{omega}),  the contribution proportional to
$e^{-m_{a_0}^2/M^2}$ becomes less dominant in Eq.(\ref{rho}),
suggesting the importance of contributions from excited states.

However, since we have calculated the OPE to the leading power
corrections only, including additional terms in the
phenomenological side becomes meaningless.   Moreover, we will
take the asymptotic value at  higher Borel masses, where the
approximation of taking the  leading power correction becomes more
reliable.   At larger Borel mass, the curve in Fig.
(\ref{rho-plot}) approach an asymptotic value for  $f_{a_0}$ at
about 20 \% of the corresponding pion value.  This is then quite
consistent with the case for the $\sigma$ case, and with the fact
that the dominant quark content of the scalar nonet is a
tetraquark.

With all our result, we have shown that a consistent picture
emerges from the QCD sum rule analysis where there is a nontrivial
coupling between the $a_0$ and the $\rho$ in the nuclear medium,
whose strength can be reliably estimated with previously
determined coupling constants. Therefore, such corrections should
always be included in estimating the $a_0$ contribution to the
dilepton spectrum from heavy ion collision\cite{Gale00,Gale01}.

\begin{table}
\centering
\begin{tabular}{|c|c|c|c|c|c|}
\hline Isospin  & $f_V$ & $g_{S}$ & $g_V$ & $\kappa $
 &   $\frac{f_s}{\frac{1}{2}\langle 0| \bar{u} i \gamma^5 u-\bar{d} i \gamma^5 d| \pi^0 \rangle}$ from present work   \\[2pt]
\hline 0 & 0.12  GeV$^2$ &  9.4  & 11.5  & 0  & 0.10
  \\[2pt]
\hline 1  & 0.12 GeV$^2$ &  2.8  &  0  & 14.2  & 0.2
  \\[2pt]
 \hline
\end{tabular}
\caption{Parameters for the $\omega-\sigma$ ($\rho-a_0(980)$)
mixing in the isospin 0 (1) channel. } \label{rho-a0}
\end{table}

\section{summary}

We have derived the QCD sum rules for the vector scalar mixing at
nuclear matter in both the isospin triplet and singlet channel.
Since the phenomenological parameters are well known except for
the overlap of the scalar interpolating field to the corresponding
ground state scalar mesons, the sum rule can be used to calculate
the value of the overlap.   We find that the overlap in both the
I=0 and 1 channels are very similar but less than 20\% of the
corresponding value in the pseudo scalar channel.  This result
confirms that both the $\sigma$ and $a_0(980)$ and probably the
remaining scalar nonets have a small quark anti-quark component and thus
suggests a large tetraquark component.  If the coupling of 
the scalar nonet
to the nucleon is very large as in the estimates in \cite{erkol,Oka},
the overlap would be even smaller.

The mixing between different spin states originates from the fact
that the nuclear medium provides a 4 momentum for the different
spin states to couple.  If the nuclear medium is not isospin
symmetric, then there could be mixing between different isospin
states also. So for example, in neutron rich matter, there could
be coupling between different isospin and spin states, such as
between the $\omega$ and the $a_0(980)$.

Experimentally, these mixing poses new challenges, as the extra
peaks from mesons with different spins could be observable from
the vector mesons emanating from the nuclear medium.  As we have
shown within the QCD sum rule analysis, a consistent picture of
mixing between scalar and vector meson emerges, whose strength can
be consistently estimated with previously determined
phenomenological couplings.   Therefore, while the $\sigma$ peak
is too wide to be observable from the $\omega$ emanating from the
nuclear medium, the $a_0$ peak will appear nontrivially in the
$\rho$ meson channel.

\section*{Acknowledgments} The work of S.H.L. has been supported  by the
Korea Research Foundation KRF-2006-C00011. T.~D. is supported by
Special Postdoctoral Research Program of RIKEN and by U.S. DOE
grant DE-FG05-84ER40154.

\begin{figure}[h]
\centering \epsfig{file=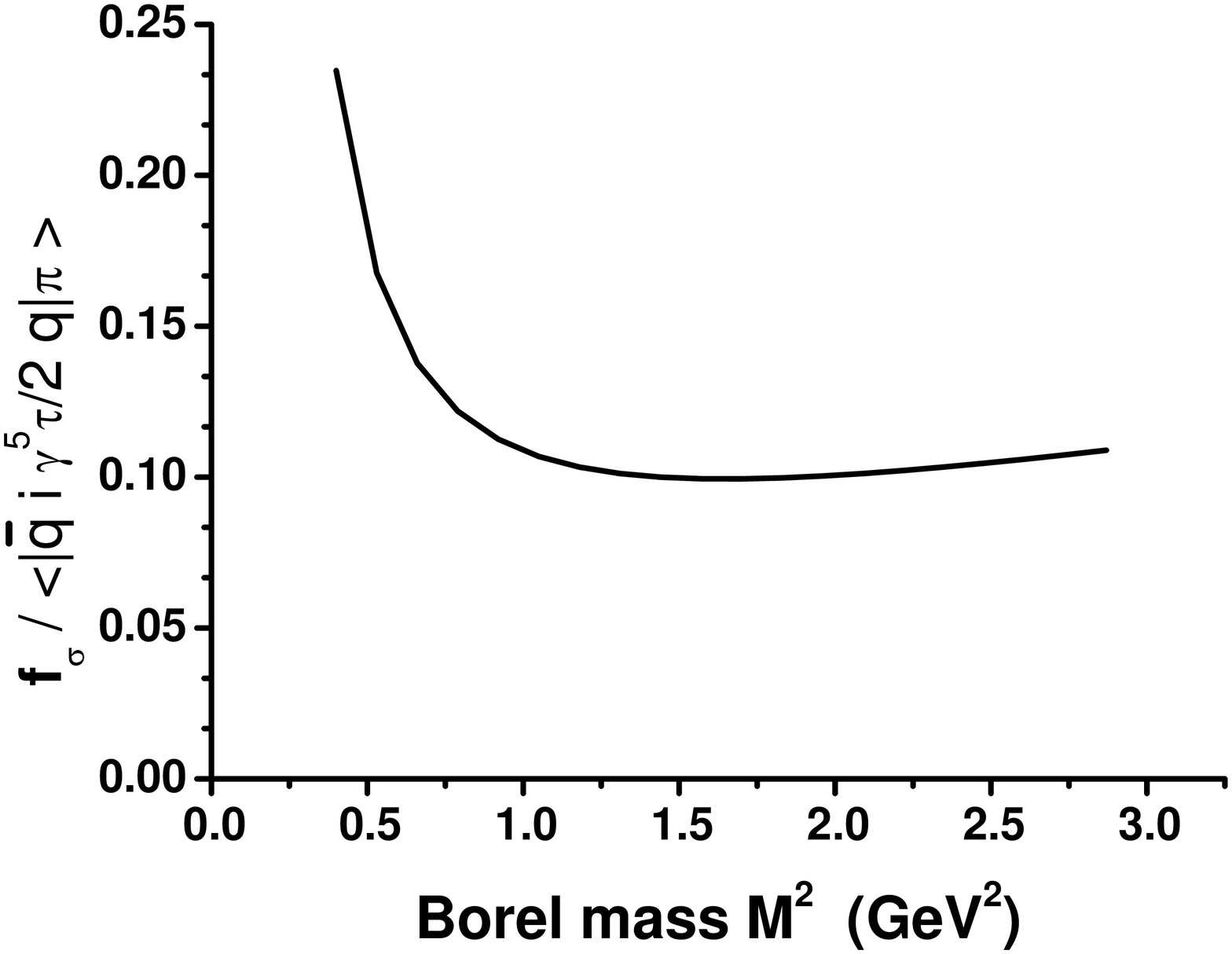, width=.8\hsize} \caption{Sum
rule for $f_\sigma=\frac{1}{2}\langle 0  |\bar{u} u + \bar{d}  d
|\sigma \rangle$ normalized to $\frac{1}{2}\langle 0 |\bar{u}i
\gamma^5 u-\bar{d} i \gamma^5 d |\pi^0 \rangle$ .}
\label{omega-plot}
\end{figure}

\begin{figure}[h]
\centering \epsfig{file=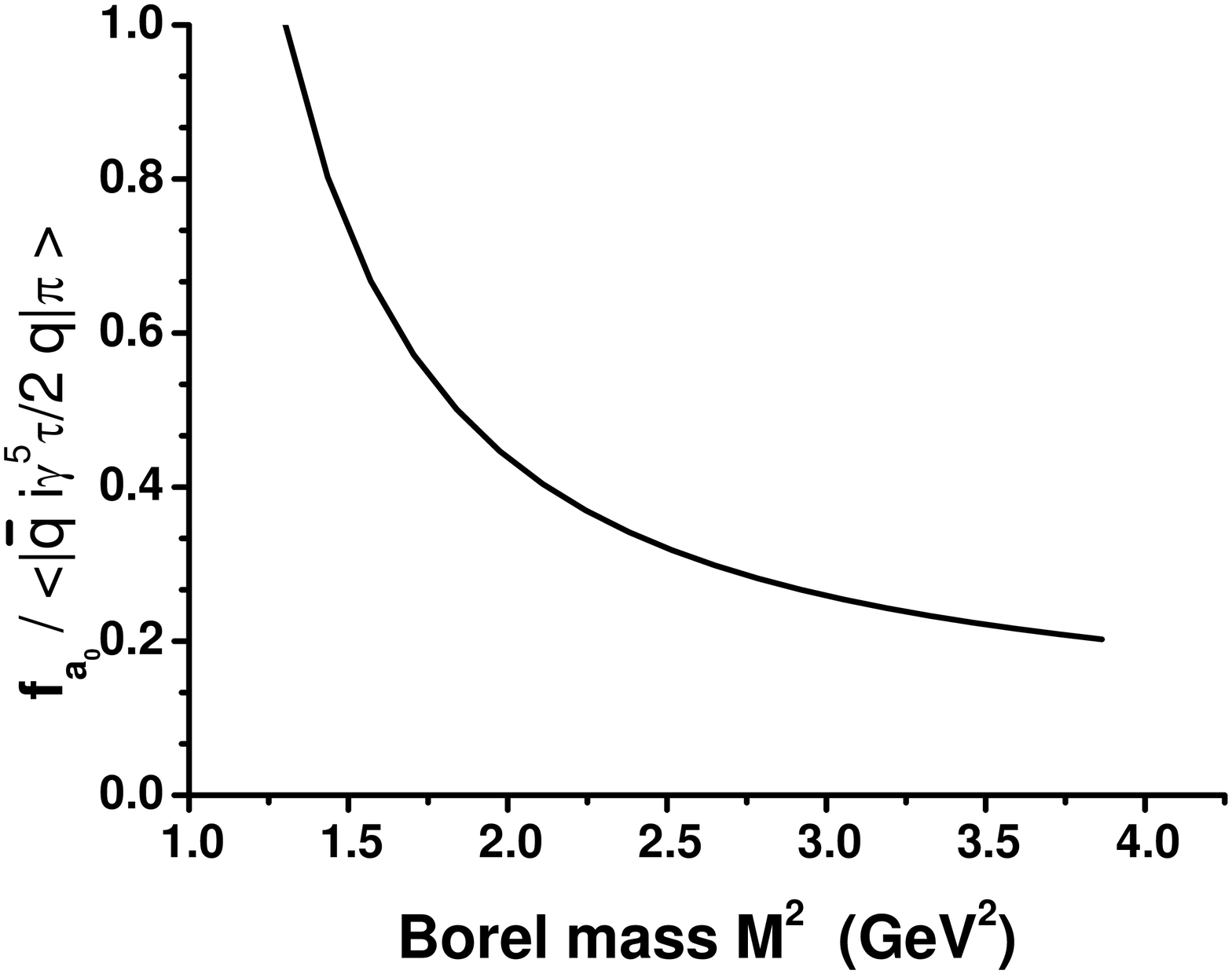, width=.8\hsize} \caption{Sum rule
for $f_{a_0}=\frac{1}{2}\langle 0  |\bar{u} u - \bar{d}  d |a_0
\rangle$ normalized to $\frac{1}{2}\langle 0 |\bar{u}i \gamma^5
u-\bar{d} i \gamma^5 d |\pi^0 \rangle$ .} \label{rho-plot}
\end{figure}

\end{document}